# ARTICLE

# Ultralow Work Function of the Electride $Sr_3CrN_3$

Cuicui Wang,[a] Miaoting Xu,[a] Keith T. Butler[b] and Lee A. Burton[a*]

Electrides have valence electrons that occupy free space in the crystal structure, making them easier to extract. This feature can be used in catalysis for important reactions that usually requires a high-temperature and high-pressure environments, such as ammonia synthesis. In this paper, we use density functional theory to investigate the behaviour of interstitial electrons of the 1-dimensional electride $Sr_3CrN_3$. We find that the bulk excess electron density persists on introduction of surface terminations, that the crystal termination perpendicular to the 1D free-electron channel is highly stable and we confirm an extremely low work function with hybrid functional methods. Our results indicate that $Sr_3CrN_3$ is a potentially important novel catalyst, with accessible, directional and extractable free electron density.

## 1. Introduction

Modern society is enabled by chemical industries. For example, it is estimated that current crop production could only sustain half of the global population without manufactured fertilisers.[1] However, this industrial process alone (the Haber-Bosch process) is estimated to consume up to 2% of the global energy supply.[2] As a result, the identification of new catalysts that can improve the efficiency of chemical processes has an enormous potential to improve quality of life *and* mitigate climate change.

Electrides are a relatively new class of ionic material that have valence electron density located in the interstitial space of the crystal structure.[3–8] The interstitial electrons are not bound to any nuclei, meaning they extract easily, *i.e.* have a low work function and diffuse easily *i.e.* have a high conductivity.[9] These properties are among those most sought after in solid catalysts, meaning electrides are promising candidates for this application.

Electrides can be considered as 0-dimensional (0D), 1-dimensional (1D) and 2-dimensional (2D), according to the degree of freedom of the anionic electron in the crystal lattice. For example, Mayenite ($Ca_{12}Al_{14}O_{32}$) can be considered 0D because the excess electrons exists in pores,[10] $Y_5Si_3$ can be considered 1D because they are in a channel,[11,12] and $Ca_2N$ can be considered 2D because they are in a plane.[13] The excess electron density for each of these materials has been observed in experiment not just predicted in theory; see refs [14] [15] [16] respectively.

The lower the dimensionality of the electrides, the more stable they are expected to be because the crystal structure helps shield the anionic electronic density from electrophilic attack. On the other hand it is expected that access to the excess electron density is relevant to catalytic performance. As a result, the 1D electrides may represent an ideal compromise between available electron density and stability.[12]

To date, there is no 1D electride in active use as a catalyst but recently Chanhom *et al.* confirmed $Sr_3CrN_3$ as a new one-dimensional electride with a partially filled d-shell transition metal.[17] The excess electron density originates from the Cr in a 4+ rather than a more typical 3+ state.[18] The structure is composed of trigonal units of $CrN_3$ planar in the a-b axis and the Sr ions form distorted 5-fold coordination environments with N anions coordinated towards 1 dimensional channels in the y-axis, in which resides the excess electron density. The crystal structure belongs to the P6$_3$/m space group (symmetry number 176), and is shown in Figure 1.

We investigate the non-polar surface terminations of $Sr_3CrN_3$ with density functional theory (DFT). We find that the interstitial electron density persists, even upon introduction of a surface termination. We also find that the surface perpendicular to the one-dimensional channel of $Sr_3CrN_3$ is the most stable despite exposing the excess electron density to the external system. Finally, we find the work function of this surface to be extremely low, which means that the electrons should be able to transfer to almost any reactants. Overall, $Sr_3CrN_3$ is an exciting new material, of interest for fundamental and applied reasons, and is deserving of further research.

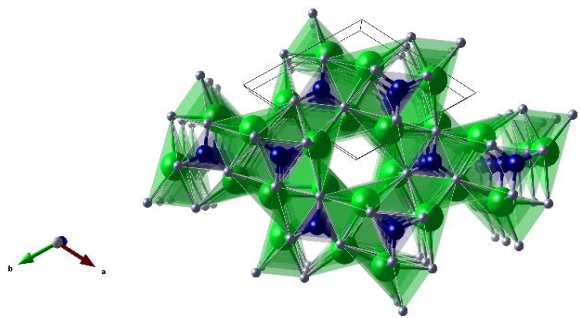

Figure 1: $Sr_3CrN_3$ crystal structure with spheres representing atoms of Sr (green), Cr (blue) and N (white). The unit cell is shown with a black box and the 1D channel is at the centre of the image.

[a.] *International Centre for Quantum and Molecular Structures, Department of Physics, Shanghai University, Shanghai 200444, China.*
[b.] *Department of Chemistry, University of Reading, Reading, RG6 6AD, UK.*
Electronic Supplementary Information (ESI) available: [details of any supplementary information available should be included here]. See DOI: 10.1039/x0xx00000x

Table 1: The lattice constants of $Sr_3CrN_3$ from DFT ionic relaxation and the values reported from experiment in the literature.[19]

| $Sr_3CrN_3$ | a | b | c | α | β | γ |
|---|---|---|---|---|---|---|
| DFT | 7.839 | 7.839 | 5.243 | 90 | 90 | 120 |
| Experiment | 7.724 | 7.724 | 5.249 | 90 | 90 | 120 |

## 2. Methods

We used density functional theory (DFT) as implemented in the Vienna Ab Initio Simulation Package (VASP) to evaluate the total the energy of compounds.[20,21] For the exchange-correlational functional, we employ a mix of Generalized Gradient Approximation (GGA) within the Perdew-Burke-Enzerhof (PBE) formulation of the exchange-correlation functional.[22] We use the Projector Augmented Wave (PAW) method for modeling core electrons with an energy cut-off of 520 eV.[23,24] This cut-off corresponds to 1.3 times the highest recommended among the pseudopotentials used. All computations are performed with spin polarization on and with magnetic ions in a high-spin ferromagnetic initialization (the system can relax to a low spin state during the DFT relaxation). We use a k-point mesh of 6x6x9, with the Monkhorst-Pack method.[25] The energy difference for ionic convergence is set to $2\times10^{-5}$ eV.

To calculate the surface energies and work function we create slab models. The method of Hinuma *et al.* is used to expand the cell to a non-polar supercell of the crystal, with a vacuum region of around 15 Å.[26] This way we obtain $Sr_3CrN_3$(001), (100), (101) and (110) slab models, as shown in Figure 2.

Surface energy, $\gamma$ is defined as the difference between surface free energy and bulk free energy.[27–29] The surface energy of each surface is calculated according to eqn (1),[30] in which, $E_{total}$ is the total energy of the constructed slab, $E_{bulk}$ is the energy the bulk material unit cell, $n$ is the number of formula units contained in the slab model and $A$ is the surface area of the slab. The factor of 2 accounts for the surfaces at either end of one slab model calculation.

$$\gamma = (E_{total} - nE_{bulk})/(2A) \quad (1)$$

To obtain an accurate work function ($\phi$) we use the hybrid functional HSE06[31–34] with a sheilding distance of $0.207\text{Å}^{-1}$ and 25% Hartree–Fock exchange. We use eqn (2)[35,36] to calculate the work function, in which ΔV is the difference in electrostatic potential between the vacuum and the macroscopic average electrostatic potential of the material (derived using the method of Butler *et al*)[36] and $E_F$ is the Fermi level calculated from the bulk, periodic material, also at HSE06 level of theory.

$$\phi = \Delta V - E_F \quad (2)$$

Table 2: The surface energy of different crystal planes of $Sr_3CrN_3$. The $Sr_3CrN_3$ (001) surface energy is found to be the smallest.

| $Sr_3CrN_3$ termination | Surface energy (J/m$^{-2}$) |
|---|---|
| 001 | 0.67 |
| 100 | 2.40 |
| 101 | 0.93 |
| 110 | 1.11 |

## 3. Results and discussion

The lattice constants obtained from relaxation of the bulk material are shown in Table 1, which is consistent with the values observed in experiment.[19] The calculated lattice vector values are within 3 % error of the experimentally reported case.

The (001), (100), (101) and (110) slab models of $Sr_3CrN_3$ are shown in Figure 2. The chosen terminations of $Sr_3CrN_3$ belong to the Tasker type II interface classification,[37] meaning that the slab is non-polar, due to the symmetrical sequence of atomic charges at the surface. As a result, the addition of a periodic surface on the crystal surface will not affect the ions inside the crystal and thereby are more likely to be stable and preserve the internal excess electron behaviour.

The surface energies of the different crystal terminations, calculated according to eqn (1),[30] are shown in Table 2. We find that Sr$_3$CrN$_3$(001) has the smallest surface energy and is therefore the most stable. When compared with the alternate surface energies of Sr$_3$CrN$_3$, e.g. (001), the relatively low energy indicates a high stability in spite of the cleavage plane bisecting the 1D electron channel. Such a relatively low surface energy is likely to have a strong impact on crystal growth and nanoparticle morphology.

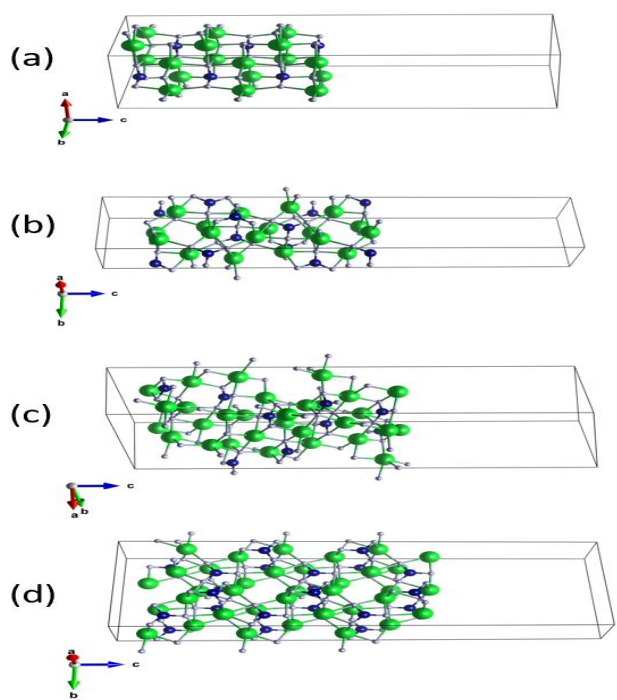

Figure 2: (a)-(d) display the constructed (001), (100), (101) and (110) surfaces of $Sr_3CrN_3$ respectively. The Sr atoms are green, Cr atoms are blue and N atoms are white. The unit cell is outlined, showing the relative size of the vacuum layer.

Figure 3 shows the valence charge density of the Sr$_3$CrN$_3$(001) slab model from the converged calculation. It can be seen that the excess electrons persists in the 1D channel, indicating that the defining feature of electrides remains even in the slab model. While the free electron density was predicted and observed experimentally for this material in bulk, this is the first time it has been observed to be robust with respect to such dramatic alteration to the crystal structure.

We continue to analyse the most stable Sr$_3$CrN$_3$(001) surface by calculating the work function. Catalysts usually work by activating chemical bonds in reactants by transferring electron density from higher energy states. Since the workfunction describes the energy required to remove an electron from the catalyst, the workfunction can therefore serve as an indicator of likely catalytic performance. [38]

$\phi$ is usually estimated by the energy difference between the Fermi level ($E_F$) and the vacuum level ($E_{vac}$).[39] Figure 4 is the calculated electrostatic potential using the slab model of Sr$_3$CrN$_3$ (001). The average electrostatic potential in the empty region corresponds with the vacuum electrostatic potential, and the difference between the this and the average electrostatic potential gives ($\Delta V$) for eqn (2). The work function of Sr$_3$CrN$_3$ (001) is calculated to be 2.14 eV.

We compare this value with the work function of other electrides from refs [40,41] as shown in Table 3. We find that even among electrides, which are known for their low work function, Sr$_3$CrN$_3$ (001) is lower still. The same theoretical method previously applied to chalcogenides finds typical values of twice or even three times the amount reported for this material.[42] Finally, we also compared this value with the work functions of elements found in experiment.[43] The Sr$_3$CrN$_3$ (001) work function is smaller than that found for all elements except for that of cesium, which is reported to be 1.95 eV from photoelectric effect experiments. Again, this highlights the relative ease of extraction for Sr$_3$CrN$_3$ compared to other materials.

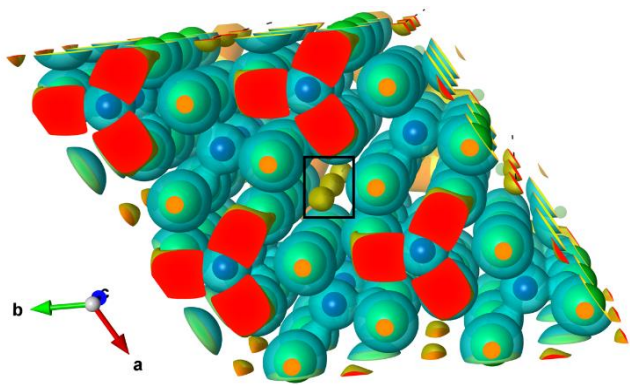

Figure 3: Valence electron density of the Sr$_3$CrN$_3$ (001) slab model. The yellow sphere enclosed by the black rectangle evinces the existence of excess electrons in the 1D channel even in the presence of a perpendicular crystal termination.

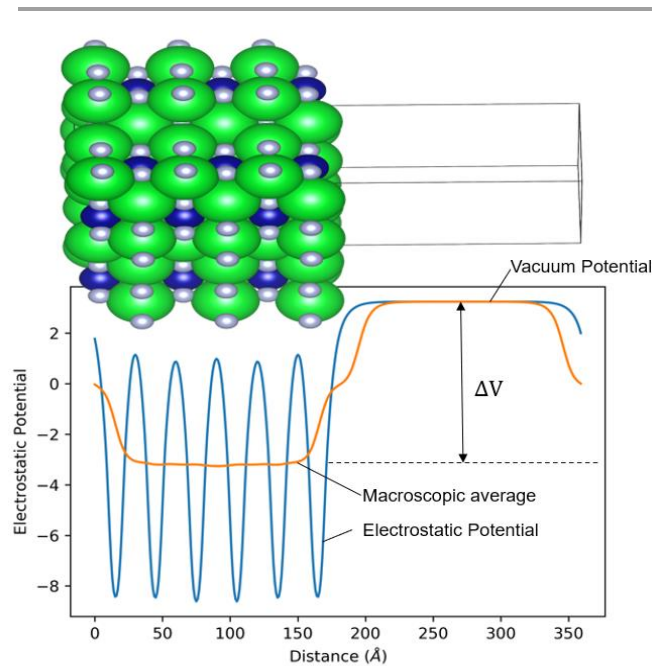

Figure 4: Slab model electrostatic potential diagram. The difference ($\Delta V$) between the average electrostatic potential and the vacuum potential is used to calculate the work function from equation (2).

In summary, we have obtained and analysed the non-polar crystal surfaces of Sr$_3$CrN$_3$. The surface that bisects the 1D electron densiy in the crystal structure is the most stable, and does not destroy the presence of the anionic electron density of the electride. Using the hybrid functional HSE06 method, it is determined that the (001) crystal termination has a very small work function of 2.14 eV, which means that the excess electron density of Sr$_3$CrN$_3$ should able to activate the chemical bonds of external reactants. What's more the uni-directionality of the electron density allows for the possibility of additional control in deploying this material as a catalyst. As a result, it is expected that the electride Sr$_3$CrN$_3$ is a unique and exciting material for catalytic applications.

Table 3: Comparison of work function of Sr$_3$CrN$_3$ (001) and work functions of other electride materials. The referenced works are included.

| Compound | Sr$_3$CrN$_3$ | C$_6$Al$_7$O$_{16}$ | Sr$_5$P$_3$ |
|---|---|---|---|
| Work function (eV) | 2.14 | 2.4 | 2.8 |
| Reference number | N/A | 41 | 42 |

## Conflict of interest

There are no conflicts to declare.